\documentclass[12pt,english,pra,preprint,showpacs]{revtex4}
\usepackage[T1]{fontenc}
\usepackage[latin9]{inputenc}
\usepackage{float}
\usepackage{amstext}
\usepackage{amssymb}
\usepackage{graphicx}
\usepackage{esint}

\makeatletter

\newcommand{\binom}[2]{{#1 \choose #2}}

\providecommand{\tabularnewline}{\\}

 \@ifundefined{textcolor}{}
 {
   \definecolor{BLACK}{gray}{0}
   \definecolor{WHITE}{gray}{1}
   \definecolor{RED}{rgb}{1,0,0}
   \definecolor{GREEN}{rgb}{0,1,0}
   \definecolor{BLUE}{rgb}{0,0,1}
   \definecolor{CYAN}{cmyk}{1,0,0,0}
   \definecolor{MAGENTA}{cmyk}{0,1,0,0}
   \definecolor{YELLOW}{cmyk}{0,0,1,0}
 }

\makeatother

\usepackage{babel}
\begin{document}

\title{Two trapped particles interacting by a finite-ranged two-body potential
in two spatial dimensions}

\author{Rostislav A. Doganov$^{1}$,  Shachar Klaiman$^{1}$,  
Ofir E. Alon$^{2}$, Alexej I. Streltsov$^{1}$, and Lorenz S. Cederbaum$^{1}$ }

\affiliation{$^{1}$Theoretische Chemie, Physikalisch-Chemisches Institut, Universit\"at
Heidelberg, D-69120 Heidelberg, Germany}

\affiliation{$^{2}$Department of Physics, University of Haifa at Oranim, Tivon
36006, Israel}
\begin{abstract}
We examine the problem of two particles confined in an isotropic harmonic
trap, which interact via a finite-ranged Gaussian-shaped potential
in two spatial dimensions. We derive an approximative transcendental
equation for the energy and study the resulting spectrum as a function
of the interparticle interaction strength. Both the attractive and
repulsive systems are analyzed. We study the impact of the potential's
range on the ground-state energy. Complementary, we also explicitly
verify by a variational treatment that in the zero-range limit the
positive delta potential in two dimensions only reproduces the non-interacting
results, if the Hilbert space in not truncated. Finally, we establish
and discuss the connection between our finite-range treatment and
regularized zero-range results from the literature.

\end{abstract}
\maketitle

\section{Introduction}

In recent years there has been an increasing interest in two dimensional
(2D) quantum systems. The condensed matter community has long been
investigating 2D quantum effects, for example, in relation to superfluid
films \cite{2D.He.Films,2D.Col.Gas}, high-temperature superconductivity
\cite{2D.High.Tc.1,2D.High.Tc.2,2D.High.Tc.3}, and low-dimensional
materials, such as graphene \cite{2D.Graphene.1,2D.Graphene.2}. The
rapid progress in atomic trapping and cooling now also allows to study
quantum systems with reduced dimensionality in the context of ultracold
trapped gases \cite{2D.Gases}. Degenerate quasi-2D Bose and Fermi
gases have already been produced in highly anisotropic ``pancake''
traps \cite{2D.BEC.Exp.1,2D.BEC.Exp.2,2D.BEC.Exp.3,2D.BEC.Exp.4,2D.BEC.Exp.5,2D.Fermi.Exp}.
This opens up the unique possibility to investigate the rich palette
of 2D quantum effects and phases in a highly controlled environment.

In dilute systems, such as the ultracold trapped gases, interactions
are usually described by a two-particle zero-range effective potential.
This approach has been particularly fruitful in one dimension, where
the delta function interaction is well behaved and there is a simple
relation between the scattering length and the interaction parameter.
In two and three dimensions, however, the delta function is not a
self-adjoint operator \cite{Div.Topics.in.Theor.Physics,Reg.2.Busch},
which gives rise to various anomalies \cite{Renorm.1,Renorm.2,Renorm.3,Div.Topics.in.Theor.Physics,J.Stat.Phys.109,Chris.Greend.Delta.3D}.
In particular, there is no scattering from a positive 2D delta function
and, for a negative one, the bound-state energy diverges. Several
approaches to overcome the arising problems and to model zero-range
interactions in 2D have been proposed in the literature: self-adjoint
extensions \cite{Self.Adjoint.1,Self.Adjoint.Solvable.Models}, renormalization
techniques \cite{Renorm.1,Renorm.2,Renorm.3}, regularization of the
delta potential \cite{Reg.1,Reg.2.Busch,van.Zyl}, and modified boundary
conditions \cite{Bound.Cond.1,Bound.2.Cond.Xea-Ji.}. 

The analytically tractable problem of two harmonically trapped particles
interacting via a zero-range potential in 2D has previously been addressed
using a regularized delta function \cite{Reg.2.Busch,van.Zyl} and
Bethe-Peierls boundary conditions \cite{Bound.2.Cond.Xea-Ji.}. These
works have obtained and studied the spectrum of the particles with
zero-range interaction as a function of the 2D scattering length \cite{Reg.2.Busch,Bound.2.Cond.Xea-Ji.}. 

In the present work we examine the problem of two harmonically trapped
particles in 2D interacting via a \emph{finite-ranged} two-body potential
modeled by a Gaussian function. We derive an approximative, yet accurate,
transcendental equation for the energy, and present the resulting
spectrum. In particular, we study the energy levels for both positive
and negative interaction and, furthermore, explore the role of the
\emph{range} of the potential on the ground-state energy. Finally,
we establish and discuss the connection between our finite-range results
and previous zero-range works. 

The article is organized as follows: in section II we derive the general
secular equation for the energy of two trapped particles interacting
via Gaussian-shaped two-body potential in 2D. In section III, 
by utilizing a variational treatment, we briefly discuss the limit of
a positive non-regularized delta potential and show that the energy
spectrum of the non-interacting system is altered only as a consequence
of truncating the Hilbert space. In Section IV A we present an efficient
high-performance approximation for the finite-range interaction, and
derive an equation for the energy of the two particles. Then, in section
IV B, we study the resulting energy spectrum and in section V compare
our finite-range findings to zero-range results from the literature.
Finally, in section VI, we summarize our results. Supplemental derivations
and numerics are deterred to the appendices.

\section{The eigenvalue equation}

We consider two particles in an isotropic harmonic trap, which are
interacting via a normalized two-body Gaussian-shaped potential. The
Hamiltonian of the system is 

\begin{equation}
H=\sum_{i=1}^{2}\left(-\frac{\hbar^{2}}{2m}\nabla_{i}^{2}\,+\,\frac{1}{2}m\omega^{2}\bar{r}_{i}^{2}\right)+\lambda_{0}V(\bar{r}_{1}-\bar{r}_{2}),\label{eq:1}
\end{equation}

\begin{equation}
V(\bar{r}_{1}-\bar{r}_{2})=\,\frac{1}{\pi s^{2}}e^{-\frac{(\bar{r}_{1}-\bar{r}_{2})^{2}}{s^{2}}}.\label{eq:2}
\end{equation}
Here $\nabla$ is the 2D Nabla operator and $\bar{r}=(x,y)$. The
problem can be separated into a non-interacting center of mass and
an interacting relative part. With the standard definitions of a reduced
mass, $\mu=m/2$, and a total mass, $M=2m$, as well as center of
mass and relative coordinates $\bar{R}=\frac{1}{2}(\bar{r}_{1}+\bar{r}_{2})$
and $\bar{r}=\bar{r}_{1}-\bar{r}_{2}$, the Hamiltonian can be rewritten
as $H=H_{cm}+H_{rel}$ with:

\begin{equation}
H_{cm}=-\frac{\hbar^{2}}{2M}\nabla_{R}^{2}\,+\,\frac{1}{2}M\omega^{2}R^{2},\label{eq:3}
\end{equation}
 
\begin{equation}
H_{rel}=-\frac{\hbar^{2}}{2\mu}\nabla_{r}^{2}\,+\,\frac{1}{2}\mu\omega^{2}r^{2}\,+\lambda_{0}V(r).\label{eq:4}
\end{equation}
 Here $H_{cm}$ is the Hamiltonian of the 2D quantum harmonic oscillator
whose solutions are well known. From now on we concentrate on the
relative part which we further write as $H_{rel}=H_{0}+\lambda_{0}V(r)$,
where $H_{0}$ reads

\begin{equation}
H_{0}=-\frac{\hbar^{2}}{2\mu}\nabla_{r}^{2}\,+\,\frac{1}{2}\mu\omega^{2}r^{2}.\label{eq:5}
\end{equation}

We start by constructing a solution $\Psi$ of the time-independent
Schr\"odinger equation from the eigenstates of $H_{0}$. We take only
eigenstates with zero angular momentum, which we denote $\varphi_{k}(r)$.
The respective energies are $\varepsilon_{k}=(2k+1)\hbar\omega$.
We note that all states with non-zero angular momentum vanish at the
origin and, therefore, should not be significantly perturbed by the
Gaussian potential if its width is sufficiently small. After substituting
the expansion $\Psi=\sum_{i=0}^{\infty}c_{k}\varphi_{k}$ into the
Schr\"odinger equation $H_{rel}\Psi=E\Psi$, and after projecting onto
a state $\varphi_{k'}(r)$, we arrive at the secular equation

\begin{equation}
c_{k'}(\varepsilon_{k'}-E)+\lambda_{0}\sum_{k=0}^{\infty}c_{k}\int\varphi_{k'}^{*}(r)V(r)\varphi_{k}(r)d\bar{r}\,=0.\label{eq:6}
\end{equation}
The integration is taken over the whole 2D plane with $d\bar{r}=2\pi rdr$.
To proceed further we need to evaluate the matrix elements appearing
in the above sum. As we show in Appendix A, this can be done analytically.
The result of the calculation is

\begin{widetext}

\begin{equation}
I_{k',k}(s)=\int\varphi_{k'}^{*}(r)V(r)\varphi_{k}(r)d\bar{r}=\frac{1}{\pi l^{2}}\left(\frac{1}{\left(\frac{s}{l}\right)^{2}+1}\right)^{k'+k+1}{\scriptstyle 2}F{\scriptstyle 1}\left(-k'\;,-k\;;\;1\;,\;\left(\frac{s}{l}\right)^{4}\;\right).\label{eq:7}
\end{equation}

\end{widetext}Here, ${\scriptstyle 2}F{\scriptstyle 1}$ is the Gauss
hypergeometric function \cite{Handbook.of.Mat.Funct.} and $l=\sqrt{\frac{\hbar}{\mu\omega}}$
is the harmonic oscillator length.

\section{Solution for contact potential\label{sub:Contact-potential}}

Before proceeding to the results for a finite-ranged interaction,
let us first examine the limit $s\rightarrow0$, in which case the
normalized Gaussian-shaped potential defined in Eq.(\ref{eq:2}) goes
into a delta function. For $s=0$ Eq.(\ref{eq:7}) takes on the form
$I_{k',k}(0)=\frac{1}{\pi l^{2}}$ and the matrix elements are independent
of the indices $k'$ and $k$. In this limit Eq.(\ref{eq:6}) reduces
to

\begin{equation}
c_{k'}(\varepsilon_{k'}-E)+\lambda_{0}\sum_{k=0}^{\infty}\frac{1}{\pi l^{2}}c_{k}=0.\label{eq:8}
\end{equation}
The sum appearing in the above equation runs over all indices $k$.
After rearranging the expansion coefficients we obtain

\begin{equation}
c_{k'}=\frac{-\lambda_{0}C}{\varepsilon_{k'}-E},\label{eq:9}
\end{equation}
where $C$ is a parameter which can depend on the energy $E$, but
is the same for all $k'$. By substituting the above expression for
the coefficients $c_{k}$ back into Eq.(\ref{eq:8}) and dividing
by $\pi l^{2}/\lambda_{0}$, we arrive at 

\begin{equation}
\frac{\hbar\omega\pi l^{2}}{\lambda_{0}}+\sum_{k=0}^{\infty}\frac{1}{2k+1-E/\hbar\omega}=0.\label{eq:10}
\end{equation}
The sum in the above equation is a general harmonic series and is
divergent \cite{Math.Methods.of.Phys.Sci}. The anomaly associated
with a contact potential in 2D stems from this diverging sum in the
current treatment. To obtain a meaningful expression we first truncate
the sum at finite $N$ and then examine the behavior for $N\rightarrow\infty$
\begin{equation}
\frac{\hbar\omega\pi l^{2}}{\lambda_{0}}+\sum_{k=0}^{N}\frac{1}{2k+1-E/\hbar\omega}=0.\label{eq:11}
\end{equation}
From the above equation we can determine the energy spectrum of the
two trapped particles interacting via a delta potential in 2D for
a given truncation $N$. 

\begin{center}
\begin{figure}[h]
\includegraphics[width=16.0cm,angle=0]{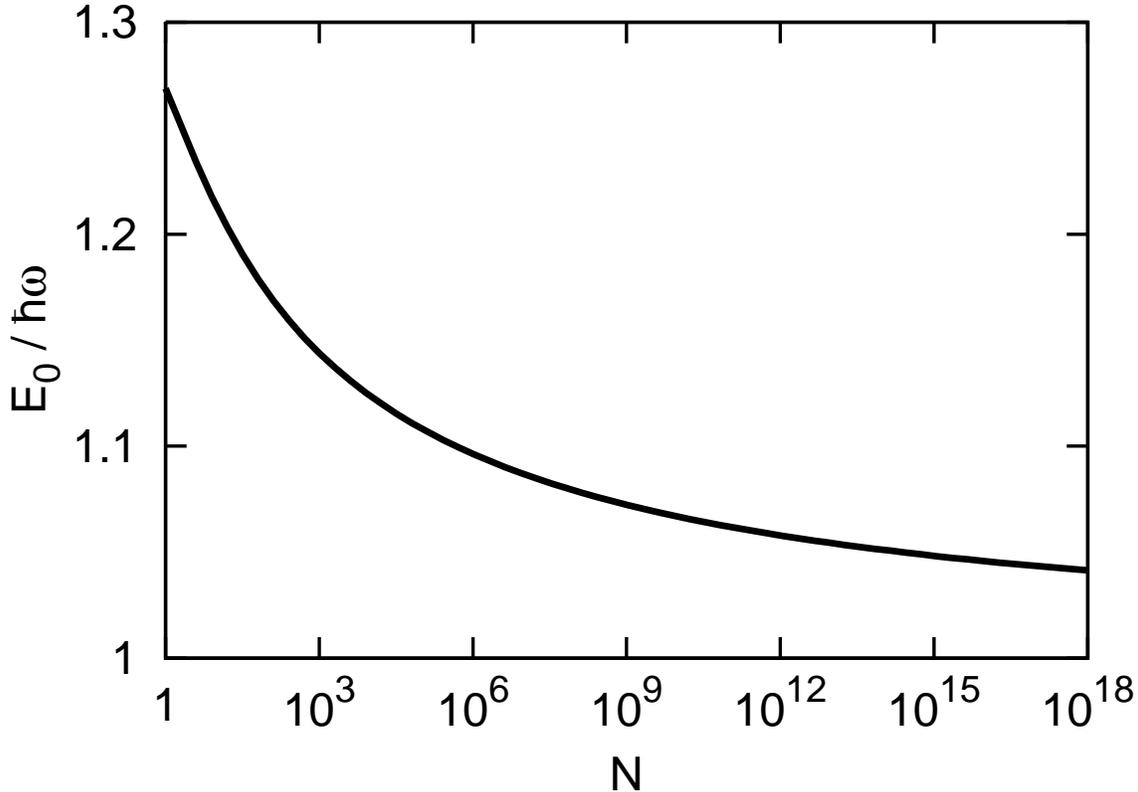}
\vglue -0.5 truecm
\caption{Ground-state energy (in the center of mass frame) of two particles
in a harmonic trap interacting via a (non-regularized) delta potential
in 2D as a function of the size of the Hilbert space. The results
are obtained by numerically solving Eq.(\ref{eq:11}) with $\lambda_{0}=1$,
$l=1$ and different $N$. Notice the logarithmic scale on the x-axis.
All quantities are dimensionless.\label{fig:Delta}}
\end{figure}
\end{center}

Let us consider the solution of Eq.(\ref{eq:11}) closest to one of
the poles appearing in the equation, say, the pole specified by $k=k'$.
We may write $E/\hbar\omega=2k'+1+\Delta k'$. This immediately leads
to

\begin{equation}
\frac{1}{\Delta k'}=\frac{\hbar\omega\pi l^{2}}{\lambda_{0}}+\sum_{k\neq k'}^{N}\frac{1}{2k+1-E/\hbar\omega}.
\end{equation}
For large $N$ the sum on the right-hand side (rhs) grows logarithmically
with $N$ for any value of $E\neq2k+1\textrm{ }(k\neq k')$, and hence
$\Delta k'$ approaches zero as $\sim1/\ln N$ . Thus, by increasing
$N$ we can make the energy levels arbitrarily close to the eigenvalues
of the respective non-interacting system. These considerations explicitly
show that in 2D the positive (non-regularized) delta potential modifies
the spectrum of two trapped particles only as a consequence of restricting
the Hilbert space. We point out that the non-interacting values are
approached logarithmically because the series in Eq.(\ref{eq:11})
diverges logarithmically. We illustrate this in Fig.(\ref{fig:Delta}),
where we plot the energy of the lowest state obtained from Eq.(\ref{eq:11})
for increasing $N$. We stress that the above considerations are rigorous
and Eq.(\ref{eq:11}) can be viewed as a variational ansatz (see Appendix
B). 

\section{Finite-range potential}

\subsection{Efficient high-performance approximation}

In order to obtain an (approximative) equation for the energy of the
two particles in the finite-range case, $s>0$, we proceed analogously
to the treatment above and make an ansatz for the expansion coefficients
$c_{k'}$. To this end we use an ansatz similar, but not identical,
to the one obtained in the Brillouin-Wigner perturbation theory (see
discussion in Appendix C)

\begin{equation}
c_{k}=\frac{-\lambda_{0}I_{0,k}(s)}{\varepsilon_{k}-E}C\label{eq:12}
\end{equation}
and obtain an equation for $E$ by substituting this expression into
Eq.(\ref{eq:6}). Since we have $I_{0,k}(0)=\frac{1}{\pi l^{2}}=const$.,
this ansatz ensures that for $s\rightarrow0$ we exactly recover the
delta potential limit of Eq.(\ref{eq:9}), which was discussed in
the previous section. We stress that the current approximation in
not variational, in contrast to the case $s=0$. However, we establish
the high accuracy of the treatment by comparing to full direct diagonalization
and zero-range results (see Appendix D and section V). 

By substituting (\ref{eq:12}) into Eq.(\ref{eq:6}) and dividing
by $\lambda_{0}$, we obtain

\begin{equation}
\frac{\hbar\omega}{\lambda_{0}}I_{0,k'}+\sum_{k=0}^{\infty}\frac{I_{0,k}(s)I_{k',k}(s)}{\varepsilon_{k}/\hbar\omega-E/\hbar\omega}=0.\label{eq:13}
\end{equation}
For each index value $k'$ the above expression gives an equation
for $E$. By setting $k'=0$ the matrix elements $I_{k',k}(s)$ take
on the form $I_{0,k}=\frac{1}{\pi l^{2}}\left(\frac{1}{\left(\frac{s}{l}\right)^{2}+1}\right)^{k+1}$
and the equation for $E$ becomes

\begin{equation}
\frac{\hbar\omega}{\lambda_{0}}I_{0,0}+\sum_{k=0}^{\infty}\frac{I_{0,k}^{2}(s)}{\varepsilon_{k}/\hbar\omega-E/\hbar\omega}=0.\label{eq:14}
\end{equation}
The series $\sum_{k=0}^{\infty}\frac{I_{0,k}^{2}}{\varepsilon_{k}/\hbar\omega-E/\hbar\omega}$
can be expressed in terms of the Lerch transcendent function $\Phi(z,s,\alpha)$
\cite{Higher.Transdent.Funct}

\begin{widetext}

\begin{equation}
\sum_{k=0}^{\infty}\frac{I_{0,k}I_{k,0}}{\varepsilon_{k}/\hbar\omega-E_{m}/\hbar\omega}=\frac{1}{(\pi l^{2})^{2}}\sum_{k=0}^{\infty}\frac{1}{2k+1-E/\hbar\omega}\left(\frac{1}{\left(\frac{s}{l}\right)^{2}+1}\right)^{2(k+1)}=\frac{\Phi\left(\frac{1}{\left(1+(s/l)^{2}\right)^{2}},1,\frac{1-E/\hbar\omega}{2}\right)}{2\pi^{2}l^{4}\left(1+(s/l)^{2}\right)^{2}}.\label{eq:15}
\end{equation}

\end{widetext}The final equation for the energy of the two trapped
particles in the center of mass frame reads 

\begin{equation}
-\frac{\Phi\left(\frac{1}{\left(1+(s/l)^{2}\right)^{2}},1,\frac{1-E/\hbar\omega}{2}\right)}{2\pi l^{2}\left(1+(s/l)^{2}\right)}=\frac{\hbar\omega}{\lambda_{0}}.\label{eq:16}
\end{equation}
Eq.(\ref{eq:16}) is the main analytical result of the paper. It allows
one to obtain the energy spectrum of two trapped particles interacting
via a Gaussian-shaped potential in 2D for given parameters $\lambda_{0}$,
$s$ and $l$.

\subsection{Energy spectrum}

\begin{center}
\begin{figure}[h]
\includegraphics[width=16.0cm,angle=0]{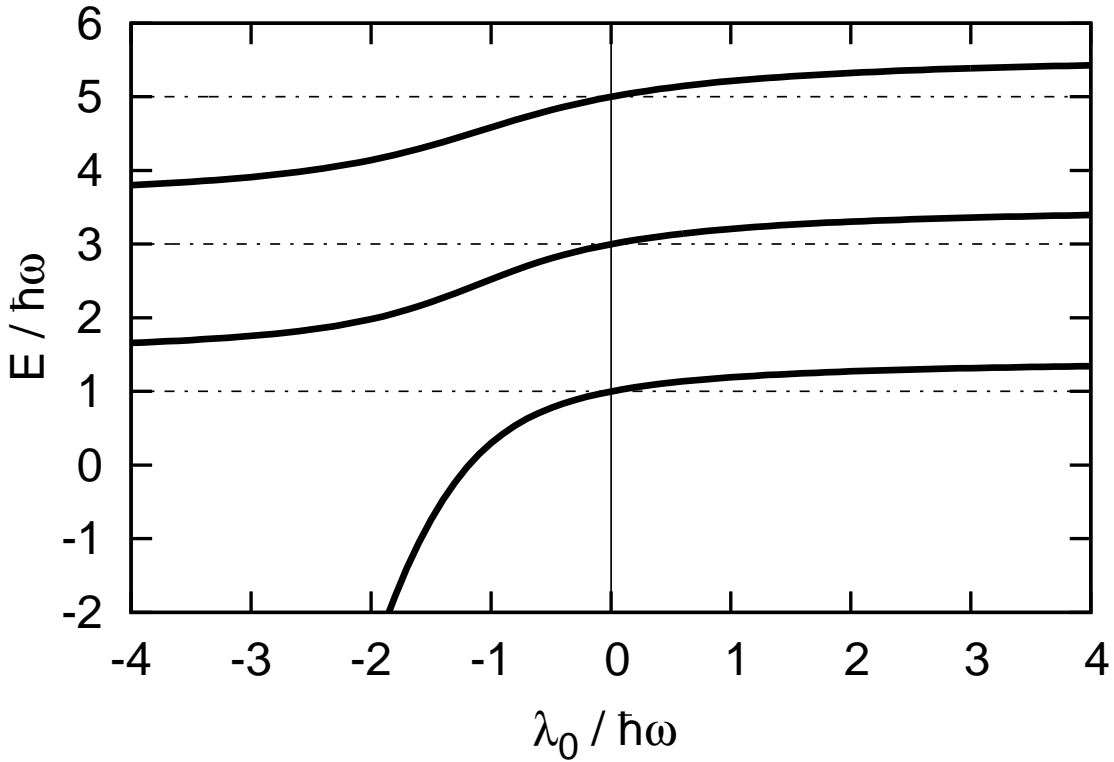}
\vglue -0.5 truecm
\caption{The energies of the first three states (in the center of mass frame)
of two trapped particles in 2D interacting via a Gaussian-shaped potential
versus the interaction strength $\lambda_{0}$. The dashed-dotted
lines show the energies of the respective non-interacting system.
Here $l=1$ and $s/l=0.1$. All quantities are dimensionless.\label{fig:Energy_vs_lambda0}}
\end{figure}
\end{center}

In this section we present the energy spectrum (in the center of mass
frame) of two trapped particles interacting via a Gaussian-shaped
potential in 2D, which results from Eq.(\ref{eq:16}). From now on
we fix the harmonic oscillator length to $l=1$. We first set $s/l=0.1$
and explore the dependence of the energy levels on the parameter $\lambda_{0}$.
The energies of the first three states versus $\lambda_{0}$ are plotted
in Fig.(\ref{fig:Energy_vs_lambda0}). For $\lambda_{0}>0$ the energies
are always above the respective non-interacting values and the system
is repulsive. As expected, at $\lambda_{0}\rightarrow0$ we recover
the non-interacting values $E/\hbar\omega=2k+1$, which correspond
to the poles of the left-hand side (lhs) of Eq.(\ref{eq:16}). For
$\lambda_{0}<0$ the particles are interacting attractively and the
ground state energy quickly diverges to $-\infty$ for $\lambda_{0}\rightarrow-\infty$.

\begin{center}
\begin{figure}[h]
\includegraphics[width=16.0cm,angle=0]{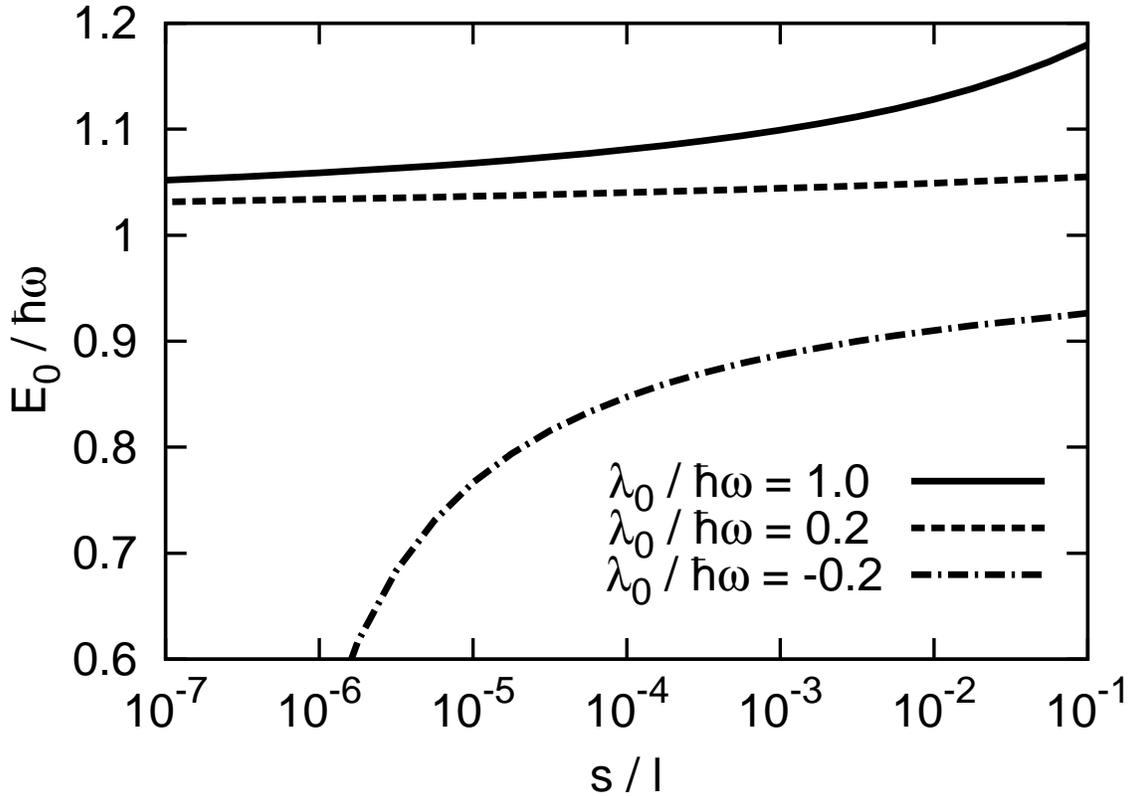}
\vglue -0.5 truecm
\caption{Energy of the ground state (in the center of mass frame) of two trapped
particles in 2D interacting via a Gaussian-shaped potential versus
the width of the interaction, $s$, for different choices of $\lambda_{0}/\hbar\omega$.
See legend inside the graph. Notice the logarithmic scale on the x-axis.
All quantities are dimensionless.\label{fig:Energy0_vs_s}}
\end{figure}
\end{center}

The Gaussian-shaped two-body potential also allows us to study the
role of the range, $s$, of the interaction. In Fig.(\ref{fig:Energy0_vs_s})
we show the dependence of the ground-state energy on $s$ for three
different values of the parameter $\lambda_{0}$. For repulsive interactions,
i.e., $\lambda_{0}>0$, we find that with decreasing $s$ the ground-state
energy logarithmically approaches the non-interacting value $\hbar\omega$.
This is in agreement with our discussion of the 2D delta potential
in section \ref{sub:Contact-potential}. We note that this behavior
is also in agreement with the formal result in \cite{Friedman}, where
it is proven that in two and more dimensions the solutions of the
Schr\"odinger equation are not affected by positive potentials with
vanishing support. For attractive interactions ($\lambda_{0}<0$)
the dependence on $s$ is more pronounced and we observe that in the
limit $s\rightarrow0$ the energy of the ground state diverges to
$-\infty$ for any negative $\lambda_{0}$. This result is also consistent
with previous studies, where it was observed that the attractive (non-regularized)
delta potential yields a bound state with an infinitely negative energy.
This is, in fact, the starting point for renormalization treatments
\cite{Renorm.3,Phys.Rev.B.Attract.Delta.Ofir}.

\section{Comparison with zero-range results}

The energy spectrum of two harmonically trapped particles with a zero-range
interaction in 2D has previously been obtained in the literature using
a regularized delta potential \cite{Reg.2.Busch}, general scattering
arguments \cite{van.Zyl}, and modified boundary conditions \cite{Bound.2.Cond.Xea-Ji.}.
In all three works the authors derive a transcendental equation for
the energy, which can be written in the form 

\begin{equation}
\tilde{\psi}\left(\frac{1-E/\hbar\omega}{2}\right)=\ln\left(\frac{l^{2}}{2a_{2D}^{2}}\right)+A.\label{eq:17}
\end{equation}
Here $\tilde{\psi}(x)$ denotes the Digamma function, $a_{2D}$ is
the 2D scattering length, and $l$ is the harmonic oscillator length.
The interaction strength is controlled through an interaction parameter
defined as $\ln\left(\frac{l^{2}}{2a_{2D}^{2}}\right)^{-1}$\cite{Reg.2.Busch,van.Zyl}.
The constant factor $A$, which is slightly different within the works
\cite{Reg.2.Busch,van.Zyl,Bound.2.Cond.Xea-Ji.}, depends on the exact
form for the effective range expansion that the respective authors
use (see the discussion at the end of \cite{van.Zyl}) and has no
consequence on the following analysis.

In order to relate our finite-range result for the energy of two trapped
interacting particles to the zero-range treatments, we employ a Taylor
expansion of the lhs of Eq.(\ref{eq:16}) around $s=0$ \cite{Mathematica}

\begin{widetext}

\begin{equation}
-\frac{\Phi\left(\frac{1}{\left(1+(s/l)^{2}\right)^{2}},1,\frac{1-E/\hbar\omega}{2}\right)}{2\pi l^{2}\left(1+(s/l)^{2}\right)}\approx\frac{1}{2\pi l^{2}}\left[\tilde{\psi}\left(\frac{1-E/\hbar\omega}{2}\right)+\ln\left(\frac{2s^{2}}{l^{2}}\right)+\gamma\right]+O[s].\label{eq:18}
\end{equation}

\end{widetext}Here $\gamma\approx0.577(2)$ is the Euler-Mascheroni
constant and $\tilde{\psi}(x)$ is again the Digamma function. By
neglecting terms of order $O[s]$ and after rearranging, we can rewrite
Eq.(\ref{eq:16}) for $s\approx0$ as 

\begin{equation}
\tilde{\psi}\left(\frac{1-E/\hbar\omega}{2}\right)=\frac{2\pi l^{2}}{\lambda_{0}/\hbar\omega}+\tilde{A}(s),\label{eq:19}
\end{equation}
with $\tilde{A}(s)=-\ln\left(\frac{2s^{2}}{l^{2}}\right)-\gamma$.
Evidently, the above equation for the energy has the same form as
the literature result in (\ref{eq:17}), and by an appropriate choice
of $\lambda_{0}$ and $s$ our ansatz reproduces the zero-range spectrum.
By comparing the rhs of Eq.(\ref{eq:19}) and (\ref{eq:17}) we immediately
see that (for weak interactions) the interaction parameter is equal
to $\frac{\lambda_{0}/\hbar\omega}{2\pi l^{2}}$ in our finite-range
analysis \cite{note_as_article}.
Thus, the interaction parameter in 2D is proportional to the factor
$\lambda_{0}$ in front of the two-body potential [see Eq.(\ref{eq:1})].

\begin{center}
\begin{figure}
\includegraphics[width=16.0cm,angle=0]{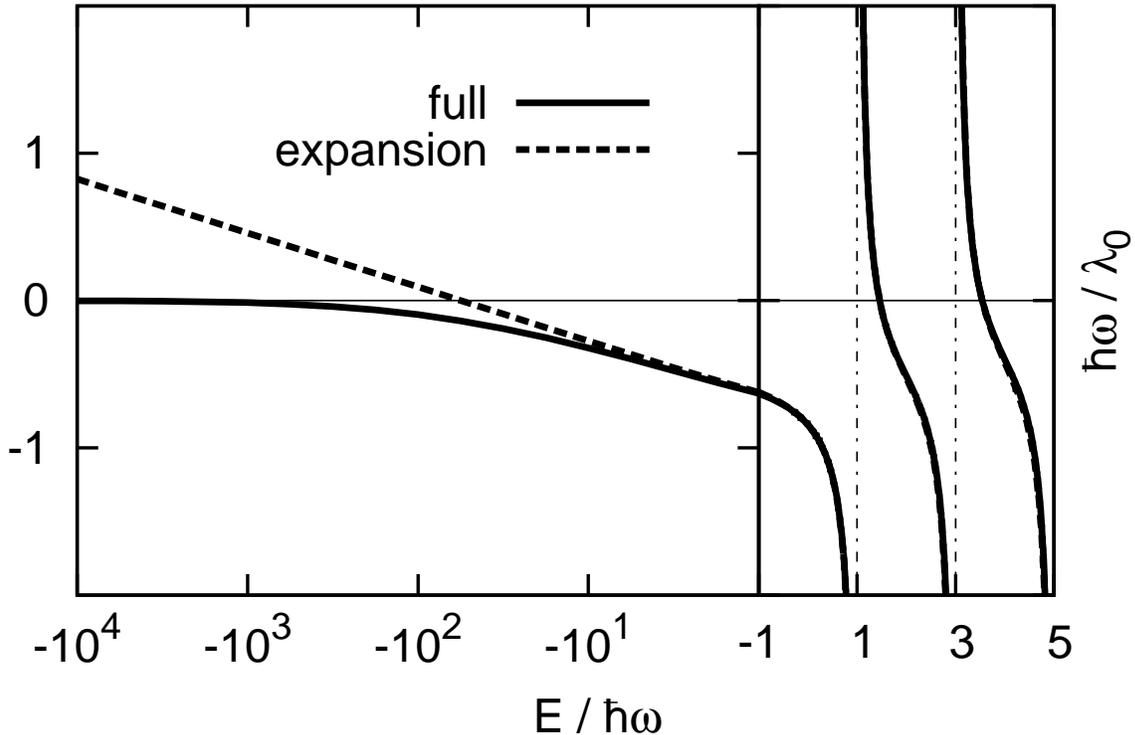}
\vglue -0.5 truecm
\caption{Plot of the lhs (thick line) and the rhs (dashed line) of Eq.(\ref{eq:18})
versus $E/\hbar\omega$ for $l=1$ and $s/l=0.1$. For $E/\hbar\omega>-1$
the two curves are practically indistinguishable. The vertical dashed-dotted
lines show the poles of both sides of Eq.(\ref{eq:18}). Notice the
logarithmic scale on the x axis. See text for discussion. All quantities
are dimensionless.\label{fig:Expansion}}
\end{figure}
\end{center}

The Taylor expansion in (\ref{eq:18}) 
leads to two significant differences between the zero-range spectrum and the finite-range spectrum 
already presented in Fig.(\ref{fig:Energy_vs_lambda0}). 
First, when the interaction is repulsive, i.e., $\lambda_0 >0$, 
Eq.(\ref{eq:19}) yields an additional deeply-bound state 
which is not present in the original finite-range spectrum 
(see also \cite{Additional.State.3D} 
for the case of hard spheres in three dimensions, 
where also a redundant state appears in the 
zero-range pseudopotential approximation).
This state appears due to the different asymptotic behavior of both
sides of (\ref{eq:18}) in the limit $E/\hbar\omega\rightarrow-\infty$.
While the lhs of (\ref{eq:18}) converges to 0, the expanded rhs diverges
to $+\infty$, see Fig.(\ref{fig:Expansion}). This also leads to
a second difference. The finite-range result yields a bound state
with an energy approaching $-\infty$ for $\lambda_{0}\rightarrow-\infty$,
while the zero-range equation yields a finite value which corresponds
to the zero crossing appearing for negative $E/\hbar\omega$ in Fig.(\ref{fig:Energy_vs_lambda0}).
This demonstrates that  the short-ranged 2D Gaussian-shaped 
interaction potential and the zero-range potential approximate 
each other well except for the ground state.

Equating the rhs of Eq.(\ref{eq:19}) and (\ref{eq:17}) allows us
to obtain a connection between the 2D scattering length, $a_{2D}$,
and the parameters of the Gaussian-shaped potential $\lambda_{0}$
and $s$. Using $A=2\ln(2)-2\gamma$ from \cite{van.Zyl}, and noticing
that the authors use the definition of the harmonic oscillator length
with the full mass, we obtain

\begin{equation}
a_{2D}\approx\sqrt{2}\, s\,\textrm{e}^{-\frac{\gamma}{2}-\frac{\pi l^{2}}{\lambda_{0}/\hbar\omega}}\label{eq:20}
\end{equation}
for the 2D scattering length of a Gaussian-shaped potential of width
$s/l\ll1$. We find good quantitative agreement when we compare the
above expression for $a_{2D}$ with numerical values from \cite{PhysRevA.79.012707}.
For $\sigma=2^{-1/2}s=0.1$ and $g=\lambda_{0}=1$ the relative difference
between our analytical estimate and the numerical value is about one
percent. Even in the regime of $\sigma=2^{-1/2}s=1$ and $g=\lambda_{0}=10$,
where the accuracy of our treatment is limited, the relative difference
in only about 8 percent.

\section{Summary}

We analyzed the problem of two harmonically trapped particles in 2D,
which interact via a finite-ranged Gaussian-shaped two-body potential.
We derived an approximative transcendental equation for the energy
which is demonstrated to be highly accurate for the ground state.
Moreover, our ansatz works well also for the excited states (see Appendix
D). Using the Gaussian-shaped potential we were able to directly study
the dependence of the ground-state energy on the range of the interaction.
We found that the effect of the short-ranged interaction on the ground-state
energy vanishes logarithmically with deceasing of the potential range,
$s$, for all positive interaction strengths. This study is complemented
by a variational treatment which shows that in the limit of a (non-regularized)
delta potential, i.e. $s\rightarrow0$, the energy spectrum of the
two (repelling) particles can be made arbitrarily close to the respective
non-interacting one by increasing the size of the Hilbert space. Finally,
we established and discussed the connection between our finite-range
result and earlier zero-range treatments reported in the literature. 

Our analysis shows that a Gaussian-shaped two-body potential 
and its zero-range pseudopotential give similar results for both repulsive and attractive interactions 
except for the lowest eigenstate of the latter. 
Here, for repulsion the zero-range pseudopotential leads to an additional dimer bound state 
which is not connected to the unperturbed system when the interaction is switched off. 
Consequently, this dimer state is difficult to reach when the interaction is smoothly switched on. 
When going
beyond two particles, analytical treatments quickly become inaccessible,
leaving one only with a numerical recourse. For the numerical many-body
simulations one usually prefers to use smooth, finite-ranged, model
interaction potentials.
Our results support the applicability of 
the Gaussian-shaped potential as a model 
of the two-body interaction in such simulations in two dimensions, 
see for example \cite{PhysRevA.79.012707}.

\begin{acknowledgments}
We thank Axel Lode and Kaspar Sakmann for very helpful discussions.
Financial support by the Deutsche Forschungsgemeinschaft (DFG) also
within the framework of the Enable fund of the excellence initiative
at Heidelberg university is acknowledged.

\bibliography{Bibil}

\begin{thebibliography}{37}
\bibitem[{not()}]{note_as_article}
\bibinfo{note}{We can also obtain this directly by substituting (\ref{eq:20})
  in $\ln\left(\frac{l^{2}}{2a_{2D}^{2}}\right)^{-1}$ and expanding for small
  $\lambda_{0}$, which leads to
  $\ln\left(\frac{l^{2}}{2a_{2D}^{2}}\right)^{-1}\approx\frac{\lambda_{0}/\hbar\omega}{2\pi
  l^{2}}$.}
\end{thebibliography}

\end{acknowledgments}
\appendix

\section{Evaluation of the matrix elements}

Here we outline the calculation of the matrix elements in Eq.(\ref{eq:7})
of the main text. The radial part of the zero angular momentum eigenfunctions
of the 2D harmonic oscillator can be written in terms of the Laguerre
polynomials $L_{k}(x)$. With $l=\sqrt{\frac{\hbar}{\mu\omega}}$
the eigenstates read \cite{Practical.Q.M.}

\begin{equation}
\varphi_{k}(r)=\sqrt{2}l^{-1}e^{-\frac{r^{2}}{2l^{2}}}L_{k}\left(\frac{r^{2}}{l^{2}}\right).\label{eq:A1}
\end{equation}
We need to evaluate the integral 

\begin{equation}
I_{k',k}=2l^{-2}\int_{0}^{\infty}e^{-\frac{r^{2}}{2l^{2}}}L_{k'}\left(\frac{r^{2}}{l^{2}}\right)\:\frac{e^{\frac{-r^{2}}{s^{2}}}}{\pi s^{2}}\: e^{-\frac{r^{2}}{2l^{2}}}L_{k}\left(\frac{r^{2}}{l^{2}}\right)r\, dr.\label{eq:A2}
\end{equation}
We start by setting $\text{\ensuremath{\Lambda}}=1+\frac{l^{2}}{s^{2}}$
and rewriting 

\begin{equation}
I_{k',k}=\frac{2l^{-2}}{\pi s^{2}}\int_{0}^{\infty}e^{-\Lambda\frac{r^{2}}{l^{2}}}L_{k'}\left(\frac{r^{2}}{l^{2}}\right)\: L_{k}\left(\frac{r^{2}}{l^{2}}\right)r\, dr.\label{eq:A3}
\end{equation}
Now we define a new variable $\rho=\frac{\Lambda}{l^{2}}r^{2}$ and
obtain 

\begin{equation}
I_{k',k}=\frac{1}{\pi\Lambda s^{2}}\int_{0}^{\infty}e^{-\rho}L_{k'}\left(\Lambda^{-1}\rho\right)\: L_{k}\left(\Lambda^{-1}\rho\right)\, d\rho.\label{eq:A4}
\end{equation}
To get rid of the $\Lambda^{-1}$ factor in the argument of $L(\Lambda^{-1}\rho)$
we use the known multiplication formula for Laguerre polynomials \cite{Handbook.of.Mat.Funct.} 

\begin{equation}
L_{k}(\Lambda x)=\sum_{i=0}^{k}\binom{k}{i}\Lambda^{i}\left(1-\Lambda\right){}^{k-i}L_{i}\left(x\right).\label{eq:A5}
\end{equation}
By substituting the above expression for $L_{k}(\Lambda^{-1}\rho)$
and $L_{k'}(\Lambda^{-1}\rho)$ into the integral of Eq.(\ref{eq:A4}),
and by making use of the orthogonality of the Laguerre polynomials
with respect to the weight function $e^{-\rho}$, we obtain

\begin{equation}
I_{k',k}=\frac{1}{\pi\Lambda s^{2}}\sum_{i=0}^{min\{k,k'\}}\binom{k}{i}\binom{k'}{i}\Lambda^{-2i}(1-\Lambda^{-1})^{k+k'-2i}.\label{eq:A6}
\end{equation}
We can write the above expression as 

\begin{equation}
I_{k',k}=\frac{1}{\pi\Lambda s^{2}}(\frac{\Lambda-1}{\Lambda})^{k+k'}\sum_{i=0}^{min\{k,k'\}}\binom{k}{i}\binom{k'}{i}\left(\frac{1}{(\Lambda-1)^{2}}\right)^{i}.\label{eq:A7}
\end{equation}
In order to express the sum in terms of the Gauss hypergeometric function
${\scriptstyle 2}F{\scriptstyle 1}(a,b;c,z)$, we rewrite the above
expression using the Pochhammer symbols $(x)_{i}$ for the rising
factorial \cite{Handbook.of.Mat.Funct.} 

\begin{equation}
I_{k',k}=\frac{1}{\pi\Lambda s^{2}}\left(\frac{\Lambda-1}{\Lambda}\right)^{k+k'}\:\sum_{i=0}^{min\{k,k'\}}\frac{(-k)_{i}(-k')_{i}}{(1)_{i}}\frac{\left(\frac{1}{(\Lambda-1)^{2}}\right)^{i}}{i!}.\label{eq:A8}
\end{equation}
We can now directly use the definition of ${\scriptstyle 2}F{\scriptstyle 1}(a,b;c,z)$
\cite{Handbook.of.Mat.Funct.} and obtain 

\begin{equation}
I_{k',k}=\frac{1}{\pi\gamma s^{2}}\left(\frac{\Lambda-1}{\Lambda}\right)^{k+k'}\:{\scriptstyle 2}F{\scriptstyle 1}\left(-k'\;,-k\;;\;1\;,\;\left(\frac{1}{\gamma-1}\right)^{2}\;\right).\label{eq:A9}
\end{equation}
Finally, by changing back to the original variables, we arrive at
Eq.(\ref{eq:7}) of the main text.

\section{Variational energy}

Here we show that the approach in Section II A for the energy spectrum
of the non-regularized delta potential is essentially a variational
treatment with $\Psi=\sum_{i=0}^{N}c_{k}\varphi_{k}$ and $c_{k}=\frac{const.}{\varepsilon_{k}-E}$.
For this purpose, let us consider $E$ just as a parameter which fulfills
Eq.(\ref{eq:11}) for a given $\lambda_{0}$ and fixed finite $N$.
The variational energy then reads

\[
\langle\Psi|H_{rel}|\Psi\rangle=\int d\bar{r}\left(\sum_{k=0}^{N}c_{k}\varphi_{k}^{*}\right)H_{rel}\left(\sum_{k'=0}^{N}c_{k}'\varphi_{k}'\right)
\]
\[
=\sum_{k=0}^{N}\frac{const.^{2}}{\varepsilon_{k}-E}\left[\frac{\varepsilon_{k}}{\varepsilon_{k}-E}+\frac{\lambda_{0}}{\pi l^{2}}\sum_{k'=0}^{N}\frac{1}{\varepsilon_{k'}-E}\right]
\]
 
\begin{equation}
=\sum_{k=0}^{N}\frac{const.^{2}}{\varepsilon_{k}-E}\left[\frac{\varepsilon_{k}}{\varepsilon_{k}-E}-1\right]=\sum_{k=0}^{N}E\frac{const.^{2}}{(\varepsilon_{k}-E)^{2}}=E.\label{eq:A10}
\end{equation}
We thus see that the variational energy is indeed equal to $E$. In
the last step we used the normalization condition for the wavefunction
$\sum_{k=0}^{N}|c_{k}|^{2}=1$.

\section{Ansatz for $c_{k}$ in the finite-range case}

Here we discuss the ansatz for the expansion coefficients in Eq.(\ref{eq:12}).
We stress that the approach is general. Let us consider a Hamilton
operator

\begin{equation}
H=H_{0}+\lambda_{0}W
\end{equation}
and assume that the eigenstates and eigenenergies of $H_{0}$ are
known. We denote them by $|\phi_{k}\rangle$ with $H_{0}|\phi_{k}\rangle=\varepsilon_{k}|\phi_{k}\rangle$.
For an eigenstate $|\Psi\rangle=\sum c_{k}|\phi_{k}\rangle$ of the
full Hamilton operator $H$ we can write

\[
(H_{0}+\lambda_{0}W)|\Psi\rangle=E|\Psi\rangle,
\]
or equivalently

\begin{equation}
(E-H_{0})|\Psi\rangle=\lambda_{0}W|\Psi\rangle.
\end{equation}
After projection on an unperturbed state $\langle\phi_{k}|$ we express
the above equation in the form

\begin{equation}
(E-\varepsilon_{k})\langle\phi_{k}|\Psi\rangle=\lambda_{0}\langle\phi_{k}|W|\Psi\rangle.
\end{equation}
From here we obtain a symbolic expression for the expansion coefficients

\begin{equation}
c_{k}=\langle\phi_{k}|\Psi\rangle=\frac{\lambda_{0}\langle\phi_{k}|W|\Psi\rangle}{(E-\varepsilon_{k})}=-\frac{\lambda_{0}\langle\phi_{k}|W|\Psi\rangle}{(\varepsilon_{k}-E)}.\label{eq:B1}
\end{equation}
Until now we have not used any approximations. Of course, the above
expression is only an implicit one, because the unknown state $|\Psi\rangle$,
which is itself dependent on $c_{k}$, appears on the rhs. Our ansatz
in Eq.(\ref{eq:12}) of the main text consists in taking only a ``first-order
approximation'' for the expansion coefficients, i.e. we substitute
$|\Psi\rangle=|\phi_{0}\rangle$ in (\ref{eq:B1}) and obtain

\begin{equation}
c_{k}=-\frac{\lambda_{0}\langle\phi_{k}|W|\phi_{0}\rangle}{(\varepsilon_{k}-E)}.\label{eq:C4}
\end{equation}
Of course, we can also substitute $|\Psi\rangle=|\phi_{k'}\rangle,\textrm{ }k'\neq0$
into Eq.(\ref{eq:C4}) and obtain similar expressions which might
be more efficient when studying excited states.

\section{Numerical comparison}

In this section we compare energies obtained by solving Eq.(\ref{eq:16})
with the numerical values determined by the full direct diagonalization
of the Hamiltonian using the matrix elements in Eq.(\ref{eq:7}).
In Tab.(\ref{tab:Table1}) we show the relative difference between
both values, i.e. $\frac{E_{DD}-E}{E_{DD}}$, where $E_{DD}$ denotes
the value obtained from direct diagonalization and $E$ denotes the
energy obtained by solving Eq.(\ref{eq:16}). The comparison is performed
for the ground state (GS) and several excited states with zero angular
momentum {[}see the first column in Tab.(\ref{tab:Table1}){]}. We
observe an excellent agreement between the values from both methods,
with $\frac{E_{DD}-E}{E_{DD}}$ in the order of $10^{-3}$ and below,
see Tab.(\ref{tab:Table1}).

\begin{center}
\begin{figure}
\includegraphics[width=16.0cm,angle=0]{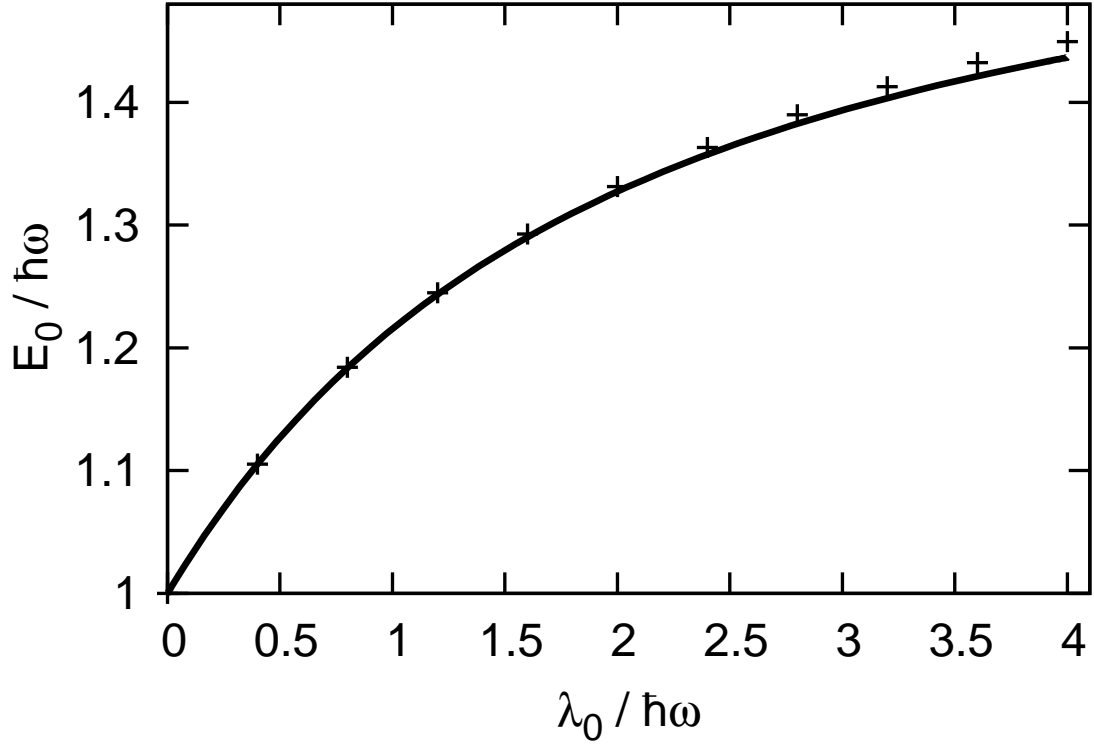}
\vglue -0.5 truecm
\caption{Comparison between the ground-state energy $E_{0}$ versus $\lambda_{0}$
obtained from Eq.(\ref{eq:16}) (thick line) and that computed by
direct diagonalization (crosses). The width is fixed to $s=0.2$ and
$l=1$. See text for more details. All quantities are dimensionless\label{fig:Comp_DD_Anal}}
\end{figure}
\end{center}

We also perform a study for a fixed width, $s=0.2$, and variable
$\lambda_{0}$. The results are shown in Fig.(\ref{fig:Comp_DD_Anal}),
where we plot the values of the ground-state energy as obtained from
Eq.(\ref{eq:16}) and direct diagonalization. The values obtained
from Eq.(\ref{eq:16}) are in excellent agreement with the ones from
direct diagonalization for $\lambda_{0}/\hbar\omega$ up to $4$.
Analogous behavior was also found for the excited states.

\begin{table}[H]
\begin{tabular}{|c|c|c|c|c|}
\hline 
 & $s=0.1$ & $s=0.2$ & $s=0.4$ & $s=0.5$\tabularnewline
\hline 
\hline 
GS & $6.1\times10^{-4}$ & $7.4\times10^{-4}$ & $6.9\times10^{-4}$ & $5.8\times10^{-4}$\tabularnewline
\hline 
1st & $2.1\times10^{-4}$ & $9.2\times10^{-5}$ & $4.9\times10^{-4}$ & $1.7\times10^{-3}$\tabularnewline
\hline 
2nd & $1.1\times10^{-4}$ & $1.6\times10^{-5}$ & $1.8\times10^{-3}$ & $4.1\times10^{-3}$\tabularnewline
\hline 
4th & $3.4\times10^{-5}$ & $1.5\times10^{-4}$ & $3.5\times10^{-3}$ & $5.2\times10^{-3}$\tabularnewline
\hline 
8th & $4.7\times10^{-6}$ & $5.7\times10^{-4}$ & $3.3\times10^{-3}$ & $3.4\times10^{-3}$\tabularnewline
\hline 
16th & $3.9\times10^{-5}$ & $1.0\times10^{-3}$ & $1.7\times10^{-3}$ & $1.4\times10^{-3}$\tabularnewline
\hline 
\end{tabular}

\caption{Numerical values of $\frac{E_{DD}-E}{E_{DD}}$ for the ground state
and several excited states with $\lambda_{0}=1.0$ and four different
choices of $s$ and $l=1$. $E_{DD}$ is the value obtained by direct
diagonalization of the Hamiltonian and $E$ is computed using Eq.(\ref{eq:16}).\label{tab:Table1}}

\end{table}

\end{document}